\documentclass[journal,draftclsnofoot,onecolumn,10pt,twoside]{IEEEtran}

\normalsize
\usepackage{cite}
\usepackage[dvips]{graphicx}
\graphicspath{{./}}
\DeclareGraphicsExtensions{.eps}

\usepackage[cmex10]{amsmath}
\usepackage{algorithmic}
\usepackage{array}

\usepackage{mdwmath}
\usepackage{mdwtab}

\usepackage{eqparbox}
\usepackage[caption=false,font=footnotesize]{subfig}

\usepackage{fixltx2e}

\ifCLASSOPTIONcaptionsoff
  \usepackage[nomarkers]{endfloat}
 \let\MYoriglatexcaption\caption
 \renewcommand{\caption}[2][\relax]{\MYoriglatexcaption[#2]{#2}}
\fi

\usepackage{url}

\begin{document}
%
% paper title
% can use linebreaks \\ within to get better formatting as desire
\title{Distributed Joint Source and Channel Coding with Low-Density Parity-Check Codes}

\author{Feng~Cen
\thanks{This work was supported by the National Natural Science Foundation of China (No. 60972035).}
\thanks{Feng Cen is with the Department of Control Science and Engineering, Tongji University, 1239 Siping Road, Shanghai, 200092, China (e-mail: feng.cen@tongji.edu.cn).}% <-this % stops a space
}

% make the title area
\maketitle
\vspace{-40pt}

\begin{abstract}
Low-density parity-check (LDPC) codes with the parity-based approach for distributed joint source channel coding (DJSCC) with decoder side information is described in this paper.
The parity-based approach is theoretical limit achievable.
Different edge degree distributions are used for source variable nodes and parity variable nodes.
Particularly, the codeword-averaged density evolution (CADE) is presented for asymmetrically correlated nonuniform sources over the asymmetric memoryless transmission channel.
Extensive simulations show that the splitting of variable nodes can improve the coding efficiency of suboptimal codes and lower the error floor.
\end{abstract}

\section{Introduction}
\IEEEPARstart{W}{e} consider applying low-density parity-check (LDPC) codes for lossless (or near lossless) distributed joint source channel coding (DJSCC) with decoder side information, which is the most basic form of lossless DJSCC.
Lossless DJSCC has been intensively studied on the basis of distributed source coding (DSC) schemes recently.
There are two dominating DSC schemes\cite{xu2007distributed}: syndrome-based approach and parity-based approach.
The former can be straightforwardly implemented with LDPC coset codes, but it is hardly used for DJSCC \cite{xu2007distributed}.
While, for the latter, it is difficult to implement efficient DJSCC with LDPC codes\cite{xu2006layered}.

Though difficult, some efforts have already been made towards efficient DJSCC using LDPC codes with the parity-based approach, such as
IRA code\cite{liveris2002joint}, LDGM code\cite{zhong2005ldgm}, systematic Raptor code\cite{xu2007distributed}, and non-systematic fountain code\cite{sejdinovic2009fountain} based schemes.
However, the regular edge degree distribution restriction of these codes on parity variable nodes (associated to parity bits) has negative impacts on the design of optimal codes.
A general LDPC code based scheme was reported in \cite{sartipi2008distributed}, but they only considered symmetrically correlated uniform sources and binary-input output-symmetric (BIOS) transmission channels and did not discuss the edge degree distribution optimization of LDPC codes.
The design of LDPC codes for DSC with the parity-based approach was also studied in the literature, such as \cite{cen2009design_commlett} and \cite{sartipi2009lossy} etc.
But they just focused on symmetrically correlated sources and cannot be easily extended to DJSCC for asymmetric transmission channels.

In this paper, we consider the parity-based approach with general LDPC codes.
We point out that the parity-based approach is a theoretical limit achievable scheme for DJSCC with decoder side information.
In order to achieve good coding performance and low error floor, different edge degree distributions are assigned to source variable nodes (associated to source bits) and parity variable nodes.
Moreover, since asymmetrically correlated nonuniform sources are more often encountered in practical applications and asymmetric physical channels are also observed in some scenarios, we present the codeword-averaged density evolution (CADE) to analyze the asymptotic performance of LDPC codes for transmitting the asymmetrically correlated nonuniform sources over the asymmetric memoryless transmission channel.

The rest of this paper is organized as follows.
Section \ref{sec:problem_statement} describes the equivalent channel coding model and shows that the parity-based approach is theoretical limit achievable.
Section \ref{sec:reg_code} presents the CADE formulas for LDPC codes.
Then, in section \ref{sec:simulation}, we show experiment results.
Finally, the conclusion is given in section \ref{sec:conclusion}.

\section{Coding scheme}\label{sec:problem_statement}
Let $X$ and $Y$ be the outputs of two correlated i.i.d random sources.
Both $X$ and $Y$ can be non-uniformly distributed.
They have a joint probability mass function $P_{XY}(x,y),x\in\{0,1\},y\in\Theta$.
Here, $\Theta$ denotes the alphabet (can be nonbinary) of $Y$ and the lower case letters $x$ and $y$ denote the realizations of their respective random variables.
Let $P_{X}(x)$ and $P_{Y}(y)$ denote the marginal probability mass functions of $X$ and $Y$, respectively.
Without loss of generality, we assume that $Y$ is taken as decoder side information.

Consider coding with the parity-based approach.
A length $k$ source sequence, $X^{k}=(X_{1},X_{2},...,X_{k})$, is encoded with a systematic LDPC code specified by its $k\times n$ generator matrix
\begin{equation}\label{equ:gen_matrix}
\mathbf{G}_{k\times n}=[\mathbf{P}_{k\times m}\;\;\mathbf{I}_{k\times k}],
\end{equation}
where $n=m+k$ and $\mathbf{I}_{k\times k}$ is an identity matrix.
The encoder generates a length $n$ codeword $\Phi^{n}=(Z^{m},X^{k})$, and then, sends the length $m$ parity bits of $\Phi^{n}$, which is formed as $Z^{m}=X^{k}\mathbf{P}_{k\times m}$, through a memoryless channel $Ch_{tr}$ to a receiver.
At the receiver side, an LDPC decoder is employed to reconstruct the codeword.
The decoder takes the sequence $\hat{Z}^{m}$ observed from $Ch_{tr}$ as the parity bits of noise corrupted codeword $\hat{\Phi}^{n}$ and $Y^{k}$ as the source bits of $\hat{\Phi}^{n}$.
By decoding $\hat{\Phi}^{n}$, the decoder outputs the source bits $\hat{X}^{k}$ of the codeword reconstruction.
Here, $\hat{X}^{k}$ is a different notation from $X^{k}$ due to the probable errors occurred in decoding.

\begin{figure}[h]
  \centering
  \includegraphics[width=0.75\columnwidth]{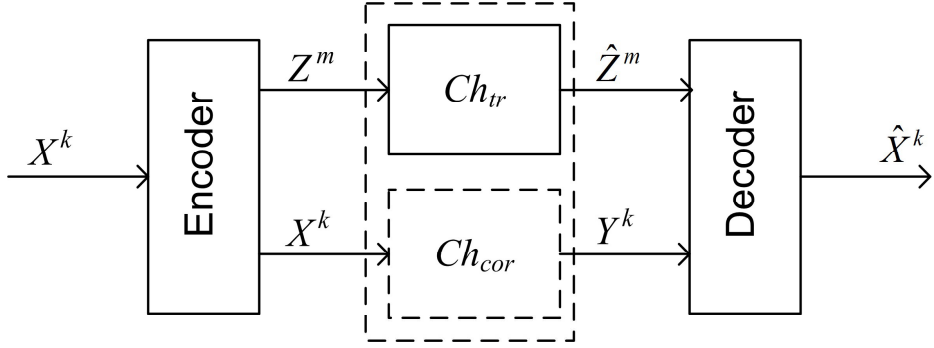}\\
  \caption{Equivalent channel coding model.}
  \label{fig:aswc_eqv_channel}
\end{figure}

From channel coding point of view, $Y^{k}$ can be regarded as the observation of a correlation channel $Ch_{cor}$ with the channel input $X^{k}$.
$Ch_{cor}$ is determined by the conditional probability mass function $P_{Y|X}(y|x)$ between $X$ and $Y$.
Therefore, the above coding approach is equivalent to channel coding for a virtual channel consisting of two parallel component channels, as shown in Fig. \ref{fig:aswc_eqv_channel}.
Although the focus of this paper is on the asymmetric $Ch_{tr}$ and the asymmetric $Ch_{cor}$, the solution in this paper is also valid for the symmetric $Ch_{tr}$ and the symmetric $Ch_{cor}$, as the symmetric channel is a special case of asymmetric channels.

Let $C_{cor}$ and $C_{tr}$ denote the channel capacities of $Ch_{cor}$ and $Ch_{tr}$, respectively, and $R_{c}:=\frac{m}{k}$ denote the coding rate of $X^{k}$ in bits.
Then, $R_{c}$ is bounded by
\setlength{\arraycolsep}{0.0em}
\begin{eqnarray}
    R_{c}&\geq&\frac{H(X)-I(X;Y)}{C_{tr}} = \frac{H(X|Y)}{C_{tr}} = R_{Th} \label{inequ:compr_rate_channel_real_low_bd}
\end{eqnarray}
\setlength{\arraycolsep}{5pt}
\!\!where $H(X)$ is the entropy of $X$, $H(X|Y)$ is the conditional entropy, $I(X;Y)$ is the mutual information, and $R_{Th}$ denotes the theoretical limit of coding rate.

\newtheorem{theorem}{Theorem}
\begin{theorem}
For DJSCC of correlated binary memoryless sources with decoder side information, $R_{Th}$ can be achieved by systematic linear block codes with the parity-based approach.
\end{theorem}
\begin{IEEEproof}
To prove the theorem, we first consider a $R_{Th}$-achieving separated source and channel coding scheme as follows.
First, $X^{k}$, $k\rightarrow \infty$, is compressed with a theoretical limit achieving LDPC coset code defined by a $k\times r$ parity check matrix $\mathbf{H}_{k\times r}$, i.e. $S^{r}=X^{k}\mathbf{H}_{k\times r}$, at the rate approaching $H(X|Y)$, i.e. $r\rightarrow kH(X|Y)$.
Then, the length $r$ sequence $S^{r}$ is encoded with a capacity-achieving LDPC code specified by an $r\times m$ generator matrix $\mathbf{G}_{r\times m}$, i.e. $Z^{m}=S^{r}\mathbf{G}_{r\times m}$, at the channel coding rate approaching $C_{tr}$, i.e. $m\rightarrow \frac{r}{C_{tr}}$.

Since block codes are utilized in both steps of the above scheme, we can combine these two encoders into just a single encoder defined by $\mathbf{\hat{G}}_{k\times m}=\mathbf{H}_{k\times r}\mathbf{G}_{r\times m}$.
Consequently, if we set $\mathbf{P}_{k\times m}$ in (\ref{equ:gen_matrix}) as $\mathbf{P}_{k\times m}=\mathbf{\hat{G}}_{k\times m}$, by employing an optimal decoder, the parity-based approach with the systematic LDPC code defined by (\ref{equ:gen_matrix}) is $R_{Th}$-achieving due to the fact that we can, at least, recover $X^{k}$ with the decoders of the above separation coding scheme.
\end{IEEEproof}

\section{Codeword-Averaged Density evolution} \label{sec:reg_code}
\subsection{Code ensemble}
Intuitively, since $Ch_{cor}$ and $Ch_{tr}$ are different channels, it is better to allow different edge degree distributions for the source variable nodes and the parity variable nodes.
Let $\mathcal{C}(\lambda_{s},\lambda_{p},\rho)$ denote the ensemble of bipartite graphs with the edge degree distributions of source variable nodes, parity variable nodes and check nodes given by $\lambda_s$, $\lambda_p$ and $\rho$, respectively.
Let $\Pi$ be the set of edges in the bipartite graph.
Let also $\Pi_{s}$ and $\Pi_{p}$ be the sets of edges incident to the source variable nodes and to the parity variable nodes, respectively.
Define $\alpha_{s} := \frac{|\Pi_{s}|}{|\Pi|}$ and $\alpha_{p} := \frac{|\Pi_{p}|}{|\Pi|}$, where $|\cdot|$ denotes the cardinality of the set.
Then, we have
\setlength{\arraycolsep}{0.0em}
\begin{eqnarray}
\lambda_{s}(x)&{=}&\sum_{i}\lambda_{si}x^{i-1}, \label{equ:lambda_s_def}\\
\lambda_{p}(x)&{=}&\sum_{i}\lambda_{pi}x^{i-1}, \label{equ:lambda_p_def}
\end{eqnarray}
\setlength{\arraycolsep}{5pt}
\!\!where $\lambda_{si}$ subject to $\sum_{i}\lambda_{si}=\alpha_{s} $ and $\lambda_{pi}$ subject to $\sum_{i}\lambda_{pi}=\alpha_{p}$ are the fraction of edges emanating from the degree $i$ source variable nodes and the fraction of edges emanating from the degree $i$ parity variable nodes, respectively.
The edge degree distribution of check nodes is defined in conventional form, i.e.
\begin{equation} \label{equ:rho_poly}
\rho(x)=\sum_{i}\rho_{i}x^{i-1},
\end{equation}
where $\rho_{i}$ subject to $\sum_{i}\rho_{i}=1$ denotes the fraction of edges emanating from the degree $i$ check nodes.

\subsection{Density evolution}
Since the all-zero codeword cannot be assumed for density evolution in asymmetric channel setting, inspired by the work of \cite{wang2005density} for conventional channel coding, we average the density over all possible codewords to analyze the average asymptotic performance of the code ensemble for DJSCC with decoder side information.

Let $m_{v_{s}c}$ or $m_{v_{p}c}$ denote the message sent out of a source variable node $v_{s}$ or a parity variable node $v_{p}$ to a check node $c$, respectively, and $m_{cv}$ denote the message passed from $c$ to to a variable node $v$ ($v$ can be either $v_{s}$ or $v_{p})$.
By $P_{s}^{(l)}(x)$ and $P_{p}^{(l)}(x)$, $x\in \{0,1\}$, we denote the codeword-averaged density of $m_{v_{s}c}$ and $m_{v_{p}c}$ at the $l$th iteration conditioned on that the corresponding variable bit takes value $x$, respectively.
We can easily write the update formulae of CADE at the variable nodes as
\setlength{\arraycolsep}{0.0em}
\begin{eqnarray}
P_{s}^{(l)}(x)&{=}&P_{s}^{(0)}(x)\otimes \lambda_{s}\left(Q^{(l-1)}(x) \right), \label{equ:src_node_de}\\
P_{p}^{(l)}(x)&{=}&P_{p}^{(0)}(x)\otimes \lambda_{p}\left(Q^{(l-1)}(x) \right), \label{equ:prty_node_de}
\end{eqnarray}
\setlength{\arraycolsep}{5pt}
\!\!where $Q^{(l)}(x), x\in \{0,1\}$ denotes the codeword-averaged density of $m_{cv}$ conditioned on that the corresponding destination variable bit takes value $x$ at the $l$th iteration.

To derive the update formula at the check nodes, we first consider the simplest case that each check node is connected to $d_{c}$ variable nodes.
Here, the edges connected to the source variable nodes and the edges connected to the parity variable nodes are not distinguished at the check node.
Hence,  the probability of the bit associated to the variable nodes taking value $x$ (denoted by $p(x)$) is the weighted average of the probability of the source bits taking value $x$ (denoted by $p_{s}(x)$) and the probability of the parity bits taking value $x$ (denoted by $p_{p}(x)$).
It can be calculated by $p(x)={{\alpha _s}{p_s}(x) + {\alpha _p}{p_p}(x)}$.
Accordingly, the density of the message sent from the variable node on a randomly selected edge should be a codeword-averaged density that can be obtained by ${P^{(l)}}(x) = {P_s^{(l)}(x) + P_p^{(l)}(x)}$.

The difference between our work and \cite{wang2005density} for the update formula at the check node  lies in that for the conventional channel coding that was considered in \cite{wang2005density}, the fraction of zeros in the typical codewords is assumed to be one-half, while, for the joint source channel coding that is considered in our work, the fraction of zeros in the typical codewords is equal to $p(0)$.
Let $\Gamma$ following from the definition in \cite{wang2005density} be the density transformation function.
Following the derivation of the equation (14) in \cite{wang2005density} and taking such a difference into account, we can derive the update formula of CADE at the check node as follows.
\setlength{\arraycolsep}{0.0em}
\allowdisplaybreaks
\begin{eqnarray}\label{equ:insertproof1}
&&Q^{(l - 1)}({x})\nonumber\\
&&= {\Gamma ^{ - 1}}\left( {\frac{{\sum\limits_{(k + x) \in \varepsilon } {{{d_c} - 1}\choose{k}} p{{(0)}^{{d_c} - 1 - k}}p{{(1)}^k}\Gamma {{\left( {{P^{(l - 1)}}(0)} \right)}^{ \otimes {d_c} - 1 - k}} \otimes \Gamma {{\left( {{P^{(l - 1)}}(1)} \right)}^{ \otimes k}}}}{{\sum\limits_{(k + x) \in \varepsilon } {{{d_c} - 1}\choose{k}} p{{(0)}^{{d_c} - 1 - k}}p{{(1)}^k}}}} \right)\nonumber\\
&&= {\Gamma ^{ - 1}}\left( {\frac{{\frac{1}{2}\left( {{{\left( {\Gamma \left( {p(0){P^{(l - 1)}}(0)} \right) + \Gamma \left( {p(1){P^{(l - 1)}}(1)} \right)} \right)}^{ \otimes ({d_c} - 1)}} + {{( - 1)}^x}{{\left( {\Gamma \left( {p(0){P^{(l - 1)}}(0)} \right) - \Gamma \left( {p(1){P^{(l - 1)}}(1)} \right)} \right)}^{ \otimes ({d_c} - 1)}}} \right)}}{{\frac{1}{2}\left( {{{\left( {p(0) + p(1)} \right)}^{{d_c} - 1}} + {{( - 1)}^x}{{\left( {p(0) - p(1)} \right)}^{{d_c} - 1}}} \right)}}} \right) \nonumber\\
&&= {\Gamma ^{ - 1}}\left( {\frac{{\Gamma {{\left( {p(0){P^{(l - 1)}}(0) + p(1){P^{(l - 1)}}(1)} \right)}^{ \otimes ({d_c} - 1)}} + {{( - 1)}^x}\Gamma {{\left( {p(0){P^{(l - 1)}}(0) - p(1){P^{(l - 1)}}(1)} \right)}^{ \otimes ({d_c} - 1)}}}}{{2(\tfrac{1}{2} + {{( - \tfrac{1}{2})}^x}{{\left( {p(0) - p(1)} \right)}^{{d_c} - 1}})}}} \right) \nonumber\\
&&= {\Gamma ^{ - 1}}\left( {\frac{{\Gamma {{\left( {p(0){P^{(l - 1)}}(0) + p(1){P^{(l - 1)}}(1)} \right)}^{ \otimes ({d_c} - 1)}} + {{( - 1)}^x}\Gamma {{\left( {p(0){P^{(l - 1)}}(0) - p(1){P^{(l - 1)}}(1)} \right)}^{ \otimes ({d_c} - 1)}}}}{{2N(x,{d_c})}}} \right),
\end{eqnarray}
\setlength{\arraycolsep}{5pt}
where $N(x,i) = \frac{1}{2} + ( - \frac{1}{2})^x \left( p(0) - p(1) \right)^{i - 1}$ is a normalization factor brought in by codeword averaging and $\mathcal{E}=\{\upsilon : 0\leq \upsilon \leq i-1, \upsilon$ is even$\}$ denotes the set of even admissible degree values.
For brevity, let us define $\langle P^{(l - 1)}\rangle := {{p(0)P^{(l - 1)}}(0) + {p(1)P^{(l - 1)}}(1)} $ and $\langle P^{(l - 1)}\rangle_{-} := {{p(0)P^{(l - 1)}}(0) - {p(1)P^{(l - 1)}}(1)}$.
Then, (\ref{equ:insertproof1}) can be simplified as
\begin{equation}\label{equ:insertproof2}
Q^{(l - 1)}({x})= {\Gamma ^{ - 1}}\left( {\frac{{\Gamma {{\left( {\left\langle {{P^{(l - 1)}}} \right\rangle } \right)}^{ \otimes ({d_c} - 1)}} + {{( - 1)}^x}\Gamma {{\left( {{{\left\langle {{P^{(l - 1)}}} \right\rangle }_ - }} \right)}^{ \otimes ({d_c} - 1)}}}}{{2N(x,{d_c})}}} \right).
\end{equation}

Regarding the general case that the edge degree distribution of check nodes follows from (\ref{equ:rho_poly}), (\ref{equ:insertproof2}) can be straightforwardly generalized as
\begin{equation}\label{equ:insertproof3}
{Q^{(l - 1)}}\left( x \right) = {\Gamma ^{ - 1}}\left( {\sum\limits_i {\frac{{{\rho _i}}}{{2N(x,i)}}} \left( {\Gamma {{\left( {\left\langle {{P^{(l - 1)}}} \right\rangle } \right)}^{ \otimes (i - 1)}} + {{( - 1)}^x}\Gamma {{\left( {{{\left\langle {{P^{(l - 1)}}} \right\rangle }_ - }} \right)}^{ \otimes (i - 1)}}} \right)} \right).
\end{equation}

\subsection{Discussion}
In fact, the equivalent channel coding model is similar to a particular setting of the channel coding for parallel channels described in \cite{pishro2005nonuniform}, i.e. only two component channels, except for the nonuniform source and asymmetric component channels.
Owing to the fact that part of the irregularity for the codes is achieved by incorporating the channel nonuniformity into the ensemble definition and more information is available in code design, we can expect that $\mathcal{C}(\lambda_{s},\lambda_{p},\rho)$ can achieve good coding performance and low error floor with fewer and smaller edge degrees.
Furthermore, by choosing both $\lambda_s$ and $\lambda_p$ equivalent, we can obtain a conventional LDPC code ensemble.
Thus, in all circumstances, the performance of the codes obtained from $\mathcal{C}(\lambda_s,\lambda_p,\rho)$ is at least as good as the codes obtained from the conventional code ensemble.

Unlike other parameters determined by application settings, the value of $p_{p}(0)$ is manually selected for CADE.
A reasonable choice is to assume $p_{p}(0)=0.5$ for CADE in practical applications, although it is possible to construct LDPC codes with $p_{p}(0)\neq 0.5$.
From the channel coding point of view, the assumption of $p_{p}(0)=0.5$ has quite small impact on the performance assessment of capacity approaching codes,
because even for asymmetric transmission channels, the mutual information for the uniformly distributed input is very close to $C_{tr}$\cite{shulman2004uniform}.
Thus, we can replace $C_{tr}$ in (\ref{inequ:compr_rate_channel_real_low_bd}) with the mutual information for the uniformly distributed input to obtain a good approximation of $R_{Th}$, which is denoted by $R_{symm}$.

\section{Simulations}\label{sec:simulation}
The binary asymmetric channel (BASC) and non-uniformly distributed binary memoryless sources were considered for simulations.
For convenience, we assume that $Y$ also takes value in $\{0,1\}$.
In general, the correlation between $X$ and $Y$ can be described by $P_{Y|X}(y=1|x=0)=\varepsilon_{01}$ and $P_{Y|X}(y=0|x=1)=\varepsilon_{10}$, which can be interpreted as the transition probabilities of a BASC as well.
Here, $\varepsilon_{01}\in [0,1]$ and $\varepsilon_{10}\in [0,1]$.

In all CADE simulations, we assume that $p_{p}(0)=0.5$.
For all finite-length code simulations, the codes with $k=80000$ are used and constructed by randomly selecting the bipartite graphs from code ensembles.
The belief propagation algorithm with a maximum iteration of $200$ is adopted for decoding and more than $1000$ codewords are transmitted for each simulation.

Irregular code $C1$ with $R_{c}=0.8$ is optimized for the setting of $p_{s}(0)=0.1$, $p_{p}(0)=0.5$, $Ch_{cor}$ with $\varepsilon_{10}=0.4$, and $Ch_{tr}$ with $\varepsilon_{01z}=0.2$ and $\varepsilon_{10z}=0.01$, where $\varepsilon_{01z}:=P_{\hat{Z}|Z}(\hat{Z}_{i}=1|Z_{i}=0)$ and $\varepsilon_{10z}:=P_{\hat{Z}|Z}(\hat{Z}_{i}=0|Z_{i}=1)$.
First, two sets of degrees, each of which consists of two degrees, with a maximum degree of $20$ are selected for the source variable nodes and the parity variable nodes.
Then, we optimize the degree distributions by using differential evolution\cite{hou2001performance} in conjunction with the CADE.
To simplify the code design, we restrict the edge degrees of check nodes to two consecutive integer values.
The code ensemble of $C1$ is given by
\setlength{\arraycolsep}{0.0em}
\begin{eqnarray}
\lambda_{s}(x)&{=}&0.2362x^{2}+0.227x^{4}, \nonumber\\
\lambda_{p}(x)&{=}&0.161x+0.3758x^{19}, \nonumber\\
\rho(x)&{=}& 0.9229x^{9}+0.0771x^{10}. \nonumber
\end{eqnarray}
\setlength{\arraycolsep}{5pt}

\begin{figure}[!t]
  \centering
  \includegraphics[width=0.88\columnwidth]{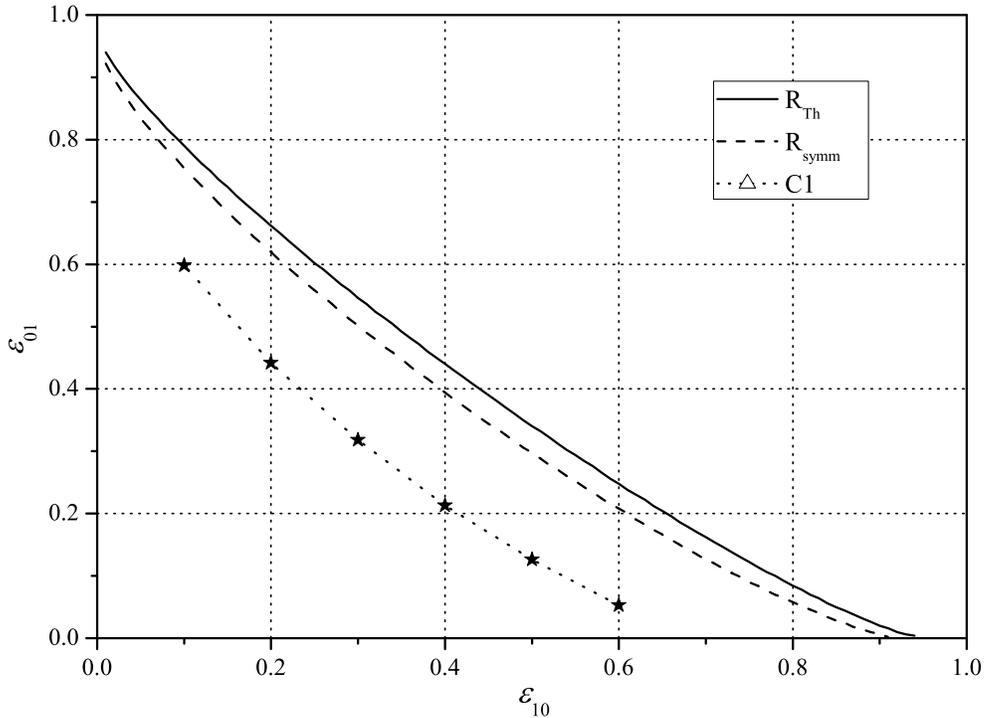}
  \caption{Theoretical limits and asymptotic thresholds of $C1$ for binary correlated nonuniform sources with $p_{s}(0)=0.1$ and $Ch_{tr}$ with $\varepsilon_{01z}=0.2$ and $\varepsilon_{10z}=0.01$.
  The maximum and minimum gaps between the asymptotic thresholds of $C1$ and the corresponding $R_{Th}$s are $0.064$ bits and $0.056$ bits, respectively.}
  \label{fig:pe_pe}
\end{figure}

From Fig. \ref{fig:pe_pe}, we can observe that $R_{symm}$ is close to $R_{Th}$ for all $\varepsilon_{01}$s.
This phenomenon demonstrates that $p_{p}(0)=0.5$ is an appropriate choice for practical applications.
Also, the similar coding performances are exhibited for various correlation settings, though $C1$ is merely optimized for one setting.
This can be attributed to the fact that each binary input asymmetric output correlation channel is equivalent to a BIOS channel\cite{chen2009equivalence} and the coding performances of LDPC codes for the BIOS channels with the same channel capacities are rather close.

\begin{figure}[!t]
  \centering
  \includegraphics[width = 0.88\columnwidth]{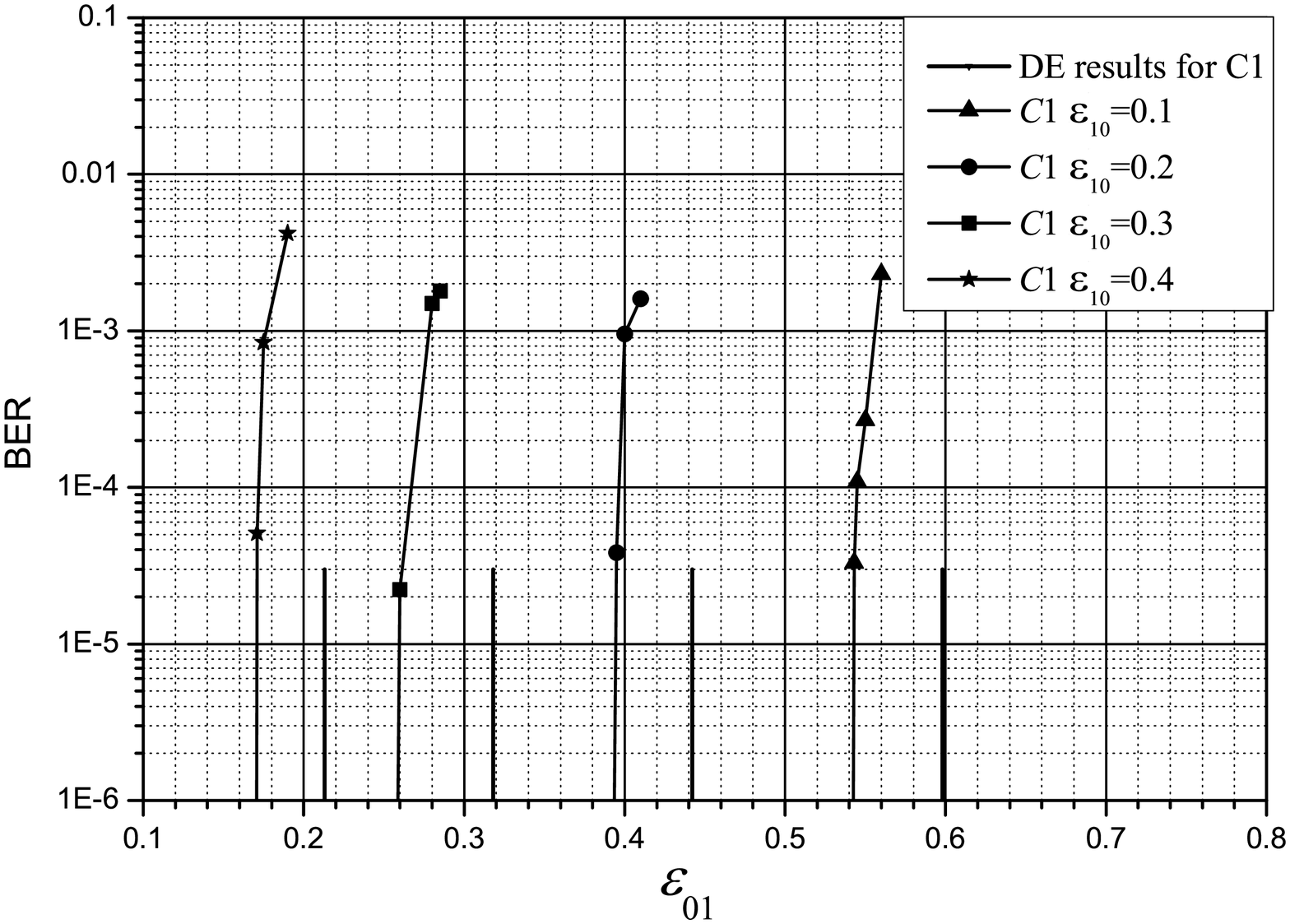}
  \caption{BER curves of the finite-length realization of $C1$ for binary correlated nonuniform sources with $p_{s}(0)=0.1$ and $Ch_{tr}$ with $\varepsilon_{01z}=0.2$ and $\varepsilon_{10z}=0.01$. The corresponding asymptotic thresholds are shown as well.}
  \label{fig:ber_finit}
\end{figure}

From Fig. \ref{fig:ber_finit}, we can observe that the waterfall portions of the BER curves of finite length codes are quite close to their respective asymptotic thresholds obtained by the CADE.
In our simulations, if the bit-error rate (BER) is less than $10^{-5}$, a commonly used criterion\cite{stankovic2006code}, the transmission is assumed to be near lossless.
The maximum gap of the near lossless thresholds of $C1$ in Fig. \ref{fig:ber_finit} from the theoretical limits is $0.08$bits and no error-floor is observed in our simulations.
Compared with the work of \cite{stankovic2006code} on asymmetric DSC, where the gaps from the theoretical limits are in between $0.121$ and $0.111$ for the source with $p_{s}\in [0.1, 0.2]$, the experiment results show that even with shorter codewords and for the asymmetric transmission channel, the better performance can be easily achieved by optimizing the edge degree distributions than by designing sophisticated coding scheme.

\begin{figure}[!t]
  \centering
  \includegraphics[width = 0.88\columnwidth]{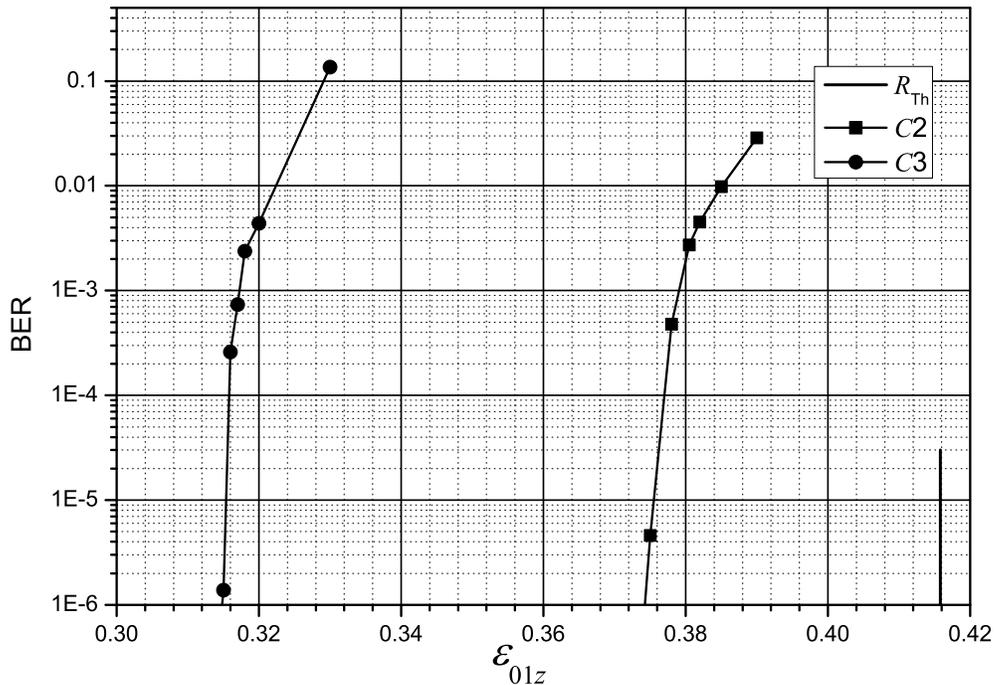}
  \caption{BER curves of the finite-length realizations of $C2$ and $C3$ for binary correlated nonuniform sources with $p_{s}(0)=0.1$, $\varepsilon_{10}=0.4$ and $\varepsilon_{01}=0.2$, and $Ch_{tr}$ with $\varepsilon_{10z}=0.01$.
  The theoretical limit is shown as well.
  The gaps between the near lossless thresholds of $C2$ and $C3$ and the theoretical limit are $0.11$ bits and $0.26$ bits, respectively.}
  \label{fig:ber_finit_r_12}
\end{figure}

To demonstrate the merit of the ensemble $\mathcal{C}(\lambda_{s},\lambda_{p},\rho)$, irregular code $C2$ with $R_{c}=1.2$ is designed for the setting of $p_{s}(0)=0.1$, $p_{p}(0)=0.5$, $Ch_{cor}$ with $\varepsilon_{10}=0.4$ and $\varepsilon_{01}=0.2$, and $Ch_{tr}$ with $\varepsilon_{10z}=0.01$.
For $C2$, a higher degree is allowed for parity variable nodes due to the worse transmission channel.
The code ensemble of $C2$ is given by
\setlength{\arraycolsep}{0.0em}
\begin{eqnarray}
\lambda_{s}(x)&{=}&0.1024x^{2}+0.3631x^{6}, \nonumber\\
\lambda_{p}(x)&{=}&0.1817x+0.0749x^{10}+0.2779x^{49}, \nonumber\\
\rho(x)&{=}& 0.2886x^{8}+0.7114x^{9}. \nonumber
\end{eqnarray}
\setlength{\arraycolsep}{5pt}
\!\!Let $C3$ be the corresponding conventional LDPC codes of $C2$.
The edge degree distribution of the variable nodes of $C3$ is given by
\setlength{\arraycolsep}{0.0em}
\begin{eqnarray}
\lambda(x)&{=}&0.1817x+0.1024x^{2}+0.3631x^{6}+0.0749x^{10} \nonumber\\
             &&+0.2779x^{49}. \nonumber
\end{eqnarray}
\setlength{\arraycolsep}{5pt}
\!\!As we can see from Fig. \ref{fig:ber_finit_r_12}, by splitting the variable nodes, it is possible to significantly improve the coding performance for suboptimal codes.

\section{Conclusion}\label{sec:conclusion}
The problem of utilizing LDPC codes for DJSCC with decoder side information for asymmetrically correlated nonuniform sources and asymmetric transmission channels is addressed in this paper.
The parity-based approach is theoretical limit achievable.
When variable nodes are split into source variable nodes and parity variable nodes, fewer and smaller edge degrees are needed for suboptimal codes and considerable gain in terms of coding efficiency compared to the conventional LDPC codes can be expected.

\ifCLASSOPTIONcaptionsoff
  \newpage
\fi

% that's all folks

% Figures

\end{document}